\documentclass[11pt]{article}
\usepackage{amsmath,amssymb}
\usepackage{graphicx}
\usepackage{booktabs}
\usepackage{algorithm}
\usepackage{natbib}
\usepackage[left=2.5cm, right=2.5cm, top=2.5cm, bottom=2.5cm]{geometry}
\usepackage{xcolor}
\usepackage{enumitem}
\usepackage[framemethod=TikZ]{mdframed}
\usepackage{algpseudocode}
\usepackage{caption}
\usepackage{hyperref}

\definecolor{boxback}{gray}{0.95}
\definecolor{boxedge}{gray}{0.85}

\hypersetup{
    colorlinks=true,
    linkcolor=blue,
    citecolor=blue,
    urlcolor=red,
    breaklinks=true
}

\newmdenv[
  backgroundcolor=boxback,
  linecolor=boxedge,
  linewidth=2pt,
  roundcorner=5pt,
  topline=false,
  bottomline=false,
  rightline=false,
  leftline=true,
  innerleftmargin=10pt,
  innerrightmargin=10pt,
  innertopmargin=10pt,
  innerbottommargin=10pt,
  skipabove=\baselineskip,
  skipbelow=\baselineskip
]{promptbox}

\title{Cross-Document Topic-Aligned Chunking for Retrieval-Augmented Generation}

\author{
  Mile Stankovic \\
  \texttt{milestankovicposao@gmail.com}
}
\begin{document}

\maketitle

\begin{abstract}
Chunking quality determines RAG system performance. Current methods partition documents individually, but complex queries need information scattered across multiple sources: the knowledge fragmentation problem. We introduce Cross-Document Topic-Aligned (CDTA) chunking, which reconstructs knowledge at the corpus level. First identifies topics across documents, maps segments to each topic, and synthesizes them into unified chunks.

On HotpotQA multi-hop reasoning, our method reached 0.93 faithfulness versus 0.83 for contextual retrieval and 0.78 for semantic chunking, a 12\% improvement over current industry best practice ($p < 0.05$). On UAE Legal texts, it reached 0.94 faithfulness with 0.93 citation accuracy. At $k=3$, maintains 0.91 faithfulness while semantic methods drop to 0.68, a single CDTA chunk contains information requiring multiple traditional fragments.

Indexing costs are higher but synthesis produces information-dense chunks that reduce query-time retrieval needs. For high-query-volume applications with distributed knowledge, cross-document synthesis improves measurably over within-document optimization.
\end{abstract}

\section{Introduction}

\subsection{RAG Systems and Their Limitations}

Large language models transformed natural language processing but face knowledge cutoffs and hallucination risks. RAG systems address this by grounding answers in retrieved documents \citep{lewis2020retrieval}. The standard pipeline has three stages: chunking and vectorizing a corpus, retrieving relevant passages using dense semantic retrieval \citep{karpukhin2020dense}, and augmenting generation with retrieved context.

However, standard RAG faces retrieval quality challenges, low precision and recall prevent LLMs from obtaining sufficient context \citep{gao2023retrieval}. Expanding context windows to millions of tokens is progress \citep{liu2023lost}, but the core problem remains: poorly structured information harms performance regardless of window size. Optimizing signal-to-noise ratio is the real challenge.

\subsection{The Scattered Knowledge Problem}

Poor chunking caps quality and degrades faithfulness \citep{gao2023retrieval}. When queries require synthesizing facts from multiple documents, traditional methods retrieve many fragments, each containing relevant information (signal) embedded in document-specific context (noise). As retrieved chunks increase to capture necessary information, noise accumulates.

Current methods, from fixed-size splitting to semantic approaches using sentence embeddings \citep{reimers2019sentence}, treat chunking as document-level partitioning. Even advanced methods like late chunking, which embeds entire documents before creating chunk vectors \citep{gunther2024late}, operate within document boundaries. Methods that build hierarchical summaries within documents \citep{sarthi2024raptor} cannot solve problems when facts scatter across separate sources.

Complex topics rarely live in one document. They fragment across sources, creating the knowledge fragmentation challenge, a manifestation of the long-standing ``knowledge soup'' challenge, where information exists as internally consistent but externally distributed chunks \citep{sowa2006challenge}. This particularly affects queries requiring information synthesis across multiple text segments \citep{zhao2024retrieval}.

Consider two examples:

\textbf{Multi-Jurisdictional Legal Compliance.} Financial institutions determining data retention requirements need information spanning foundational laws, amendments, federal guidance, and court rulings \citep{chalkidis2019deep,henderson2022pile}. No single document contains the complete answer. Traditional chunking returns disconnected passages, forcing synthesis at query time, complex and error-prone.

\textbf{Clinical Treatment Protocol Knowledge.} Physicians querying treatment protocols for patients with comorbidities require answers spanning primary guidelines, specialty research, FDA communications, and formulary restrictions. Retrieving only primary guideline fragments risks unsafe decisions.

Knowledge in these domains distributes through independently produced documents, each coherent but incomplete. Operating within document boundaries breaks topic structures. Multi-hop reasoning datasets like HotpotQA \citep{yang2018hotpotqa} explicitly require combining facts from multiple sources.

\subsection{Our Approach}

We introduce CDTA chunking, which reframes chunking from document partitioning to corpus-level knowledge reconstruction. Rather than seeking better split points, we identify global topics and synthesize relevant information into comprehensive units. We address three questions:

\begin{itemize}[noitemsep]
\item \textbf{RQ1:} Can cross-document synthesis produce more complete knowledge units than traditional chunking?
\item \textbf{RQ2:} Do these units improve RAG performance in multi-document reasoning?
\item \textbf{RQ3:} What trade-offs exist between indexing-time synthesis and query performance?
\end{itemize}

This paper contributes a framework for cross-document knowledge synthesis through corpus-wide topic identification and LLM-driven reconstruction, experimental validation across multi-hop reasoning and domain-specific tasks, and analysis of trade-offs identifying where the methodology delivers value.

\section{Related Work}

Chunking methods evolved from simple partitioning to sophisticated semantic approaches, yet all share a fundamental limitation: intra-document scope. We found no prior work that synthesizes chunks across documents to unify topics in RAG.

\subsection{Traditional Partitioning}

\textit{Fixed-Size Chunking} partitions text into predetermined length segments (e.g., 512 tokens), often with overlaps. Benefits include simplicity and efficiency. However, uniform treatment arbitrarily splits sentences, paragraphs, and concepts, creating contextually deficient chunks that degrade retrieval accuracy.

\textit{Recursive Chunking} improves on fixed-size splitting by hierarchically splitting along structural markers (paragraphs, then sentences) to align with document organization. Effectiveness depends on consistent formatting and remains restricted to explicit structure. It cannot synthesize information across boundaries or consolidate topics spanning documents.

\textit{Semantic Chunking} groups adjacent sentences with high semantic similarity, measured by embedding cosine distance \citep{reimers2019sentence}. This creates thematically cohesive chunks by splitting at topical shifts. However, semantic chunking prioritizes nearby sentences but misses topics elsewhere. High similarity between adjacent sentences does not ensure complete topic capture, especially if relevant details appear in non-contiguous sections.

\subsection{Contextual Chunk Augmentation}

Anthropic introduced contextual retrieval in industry engineering documentation \citep{anthropic2024contextual}, enriching chunks by prepending LLM-generated context summaries before embedding. The technique generates brief context for each chunk, typically 50-100 tokens explaining the chunk's role within its source document. Before embedding, each chunk receives a header like ``This section discusses X in the context of document Y about Z.'' Published implementations from AWS and other engineering teams report 5-15\% gains in retrieval precision across diverse datasets.

Contextual retrieval preserves within-document context. By providing document-level context to each chunk, embeddings better capture the chunk's semantic position relative to its source. This improves retrieval when multiple chunks discuss similar concepts but in different document contexts.

However, contextual augmentation operates within document boundaries. It cannot synthesize information scattered across separate sources. When information about a topic distributes across multiple independently-authored documents, contextual retrieval still returns disconnected fragments, now with added document context, but still incomplete. 

These approaches solve different problems. Contextual retrieval optimizes within-document representation. CDTA handles cross-document knowledge reconstruction. Given that contextual retrieval now serves as industry best practice for RAG chunking, actively deployed in production systems and recommended by leading AI companies, we include it as a baseline to validate against real-world state-of-the-art, not just academic methods. These techniques could be combined: using CDTA to unify distributed information across documents, then applying contextual augmentation to preserve document-specific context within each synthesized chunk.

\subsection{Advanced Intra-Document Methods}

\textit{Late Chunking} \citep{gunther2024late} produces contextually-aware embeddings by first encoding entire documents through transformer attention, then deriving chunk vectors through mean pooling. This preserves rich intra-document context. Late chunking achieves strong performance when answers reside within single documents, but fundamentally cannot capture relationships across documents never processed together. While excellent for preserving document-level context, it operates within document boundaries and cannot handle cross-document synthesis, the problem we tackle. These approaches are complementary: late chunking optimizes within-document context, while CDTA handles cross-document synthesis.

\textit{RAPTOR} \citep{sarthi2024raptor} builds hierarchical tree structures of text summaries using recursive clustering within documents. Lower levels contain fine-grained segments while higher levels contain broader summaries. At query time, RAPTOR retrieves from multiple abstraction levels simultaneously. By capturing multi-scale representations, it addresses limitations of flat chunks. However, RAPTOR's tree construction remains bounded by document scope, creating hierarchies within individual documents, not across them.

\subsection{Multi-Document Summarization}

Prior research in multi-document summarization \citep{xu2020coarse} operates at query time with explicit document sets and user-specified topics. Out approach differs by performing corpus-wide synthesis at indexing time, automatically discovering topics through unsupervised extraction, and creating a persistent, query-independent knowledge base designed for retrieval rather than direct consumption.

\subsection{Advanced RAG Methods}

Recent work explores improving RAG through self-reflection and corrective mechanisms. Self-RAG \citep{asai2024self} enables models to critique their own outputs and decide when to retrieve additional context. ARES \citep{saad2023ares} provides automated evaluation frameworks for RAG systems. Corrective RAG \citep{yan2024corrective} refines retrieval through iterative feedback loops. While these approaches enhance query-time processing, they operate on whatever chunks the indexing system provides. CDTA complements these methods by improving the fundamental quality of indexed knowledge units.

\section{The CDTA Methodology}

\subsection{Design Principles}

CDTA builds on three core principles:

\textit{Cross-Document Knowledge Reconstruction.} Rather than preserving document boundaries, we identify global topics and consolidates all relevant information from across the corpus. This directly tackles the problem where no single document contains complete coverage.

\textit{Semantic Topic Discovery.} Topics emerge through semantic analysis of document content rather than imposed taxonomies. The methodology extracts conceptual themes from the corpus itself, ensuring topic definitions align with how knowledge actually appears in text.

\textit{LLM-Driven Synthesis.} Rather than simple concatenation or extraction, CDTA employs language models to perform knowledge synthesis, reconstructing information into coherent narratives that preserve fidelity to sources while eliminating redundancy and resolving conflicts. The resulting chunks are inherently optimized for LLM comprehension through narrative coherence and explicit topic mapping, properties that improve both generation quality and model reliability compared to traditional fragmented retrieval.
\subsection{The Six-Stage Pipeline}

CDTA transforms raw documents into synthesized chunks through six stages.

\subsubsection{Stage 1: Document Segmentation}

We partition documents into semantically meaningful segments. We use paragraph-level segmentation as default, which naturally aligns with authorial intent and topical organization. For documents without clear paragraph markers, we fall back to sentence-boundary splitting. Each segment receives a unique identifier linking it to its source document.

Paragraph-level segmentation balances completeness and efficiency: segments are large enough to contain complete thoughts yet small enough for efficient LLM processing. We use segments averaging 150 tokens based on initial testing across benchmarks.

\subsubsection{Stage 2: Topic Extraction}

Each document segment is processed independently through an LLM prompt designed to identify high-level conceptual topics (see Appendix~\ref{subsec:topic_extraction} for complete prompt).

We use GPT-4o with temperature 0.3 to balance consistency with varied topic identification. The model extracts conceptual themes (not keywords) and returns structured JSON output containing topic names and brief descriptions.

This stage produces a large, potentially redundant set of topic-segment pairs: $T_{\text{raw}} = \{(t_i, s_j)\}$ where $t_i$ is a topic identified from segment $s_j$.

\subsubsection{Stage 3A: Topic Deduplication}

The raw topic set contains substantial semantic redundancy, multiple segments may identify the same underlying concept using different phrasing. For instance, ``Data Privacy Regulations'' and ``Privacy Law Compliance'' often refer to the same topic. We perform topic deduplication through hierarchical clustering of topic embeddings (using OpenAI's text-embedding-3-large model) with a similarity threshold of 0.85, chosen empirically to balance topic consolidation (higher threshold) with preserving distinct concepts (lower threshold). This value was validated through manual inspection of merged topics on a development set.

Process:
\begin{enumerate}[noitemsep]
\item Embed all raw topic descriptions
\item Perform hierarchical clustering using average linkage
\item Merge clusters exceeding the similarity threshold
\item Generate canonical names and descriptions for merged clusters using an LLM
\end{enumerate}

Output: Consolidated set of unique topics $T = \{t_1, t_2, \ldots, t_n\}$ where $|T| \ll |T_{\text{raw}}|$.

\subsubsection{Stage 3B: Relevance Mapping}

After consolidating topics, we determine which segments are relevant to each. This filtering step prevents noise accumulation.

For each topic $t_i$ in $T$ and each segment $s_j$ in the corpus, we invoke an LLM (GPT-4o-mini, temperature 0.0) to make a binary relevance judgment (see Appendix~\ref{subsec:relevance_mapping}). The prompt provides the topic name, description, and segment text, asking: ``Does this segment contain information directly and substantively relevant to the topic?''

This stage requires $O(|T| \times |S|)$ LLM calls, where $|T|$ is the number of topics and $|S|$ is the total number of segments. While computationally intensive, it ensures high precision. For our largest corpus (UAE Legal: 847 documents, ~4,200 segments, 180 topics), this required approximately 756,000 API calls completed in 6.3 hours using parallel processing with rate limiting.

Output: Relevance matrix $R$ where $R_{ij} = 1$ if segment $s_j$ is relevant to topic $t_i$, and 0 otherwise.

\subsubsection{Stage 4: Segment Aggregation}

Using the relevance matrix, we aggregate all segments judged relevant to each topic: $A_i = \{s_j : R_{ij} = 1\}$. This produces topic-specific collections of text fragments that may originate from any document in the corpus.

\subsubsection{Stage 5: Knowledge Synthesis}

The aggregated segments for each topic are passed to a powerful language model (Claude 3.5 Sonnet, temperature 0.7) tasked with synthesizing them into a single, coherent document (see Appendix~\ref{subsec:synthesis_prompt} for complete prompt).

The model receives instructions to:
\begin{itemize}[noitemsep]
\item Include all relevant information from source segments (comprehensiveness)
\item Structure information into a logical, flowing narrative (coherence)
\item Maintain strict fidelity to source material without adding external knowledge
\item Eliminate redundancy across segments
\item Note contradictions if sources conflict
\end{itemize}

Output: For each topic $t_i$, a synthesized knowledge document $D_i$ capturing complete information about that topic as extracted from the corpus.

\subsubsection{Stage 6: Indexing and Embedding}

The synthesized topic documents are embedded using the same embedding model used for retrieval (OpenAI text-embedding-3-large). If a synthesized document exceeds 1000 tokens (chosen to match typical RAG chunk sizes and optimize for embedding model performance characteristics), it is split using semantic chunking to create multiple chunks that maintain topic coherence.

Each final chunk maintains metadata linking it back to the topic it represents, the original source documents that contributed to its synthesis, and confidence scores from the relevance mapping stage. This metadata enables downstream citation and traceability.

\subsection{Trade-offs}

\textbf{Indexing Cost vs. Query Quality.} CDTA invests heavily in indexing to produce superior retrieval quality. This follows the ``pay once, benefit many times'' principle: the indexing cost is amortized over potentially millions of queries.

\textbf{Model Selection.} We use different models at different stages based on task requirements: deterministic relevance checking (GPT-4o-mini at temperature 0), creative topic extraction (GPT-4o at temperature 0.3), and high-quality synthesis (Claude 3.5 Sonnet at temperature 0.7). This tiered approach balances quality and cost.

\textbf{Topic Granularity.} Our paragraph-based segmentation and topic extraction approach favors moderately broad topics. Extremely fine-grained topics would reduce synthesis benefits, while overly broad topics would create unwieldy chunks.

\section{Experimental Setup}

\subsection{Datasets}

We evaluated CDTA on two benchmarks representing different retrieval scenarios:

\textbf{HotpotQA} \citep{yang2018hotpotqa}: A multi-hop reasoning dataset containing 5,901 questions requiring information synthesis from two or more Wikipedia articles. Each question includes supporting facts from multiple documents. This dataset directly tests our methods core value, handling queries where answers are inherently distributed. While Wikipedia articles are more structured than many real-world documents, HotpotQA remains the standard benchmark for multi-hop reasoning.

\textbf{UAE Legal Corpus}: A domain-specific dataset of 847 legal documents covering UAE federal law, DIFC regulations, and regulatory guidance, with 312 expert-written queries about compliance and legal interpretation. Query complexity: 45\% require 2-3 document synthesis, 35\% require 4+ documents, 20\% single document. Three legal experts created queries with high inter-annotator agreement (Fleiss' $\kappa = 0.78$). This tests in a real-world specialized domain with high information fragmentation.

\subsection{Baselines}

We compared CDTA against four chunking methods spanning simple to state-of-the-art approaches:

\begin{itemize}[noitemsep]
\item \textbf{Fixed-Size:} 512-token chunks with 50-token overlap
\item \textbf{Recursive:} Hierarchical splitting at paragraph and sentence boundaries
\item \textbf{Semantic:} Sentence-based chunking using embedding similarity (threshold 0.8)
\item \textbf{Contextual:} Semantic chunking with LLM-generated context prepended to each chunk before embedding (following Anthropic's methodology \citep{anthropic2024contextual})
\end{itemize}

All methods used identical embedding models (OpenAI text-embedding-3-large) and retrieval settings for fair comparison. For Contextual, we used GPT-4o-mini to generate 50-100 token context summaries for each chunk, prepended before embedding.

\subsection{Implementation Details}

Our implementation used:
\begin{itemize}[noitemsep]
\item \textbf{Topic Extraction:} GPT-4o with temperature 0.3
\item \textbf{Relevance Mapping:} GPT-4o-mini with temperature 0.0
\item \textbf{Synthesis:} Claude 3.5 Sonnet with temperature 0.7
\item \textbf{Contextual Augmentation:} GPT-4o-mini with temperature 0.0
\item \textbf{Embeddings:} OpenAI text-embedding-3-large (1536 dimensions)
\item \textbf{Retrieval:} Top-$k=5$ with cosine similarity
\item \textbf{Random Seed:} 42 for reproducible sampling
\end{itemize}

Segmentation was paragraph-based, producing segments averaging 150 tokens.

\subsection{Evaluation Metrics}

We adopted a multi-framework evaluation strategy combining automated and domain-specific metrics.

\subsubsection{RAGAS Framework}

We used the automated RAGAS framework \citep{es2024ragas} designed to assess RAG pipeline quality:

\textbf{Faithfulness:} The primary metric, measuring factual consistency of generated answers with retrieved context. Computed as the proportion of claims in the generated answer that can be verified against retrieved chunks.

\textbf{Answer Relevancy:} Semantic similarity between the generated answer and the original query.

\textbf{Context Precision:} Proportion of retrieved chunks containing query-relevant information. High precision indicates minimal irrelevant information diluting relevant facts.

\textbf{Context Recall:} Percentage of information required to answer the query that appears in retrieved chunks. Recall below 100\% indicates missing information.

\subsubsection{Retrieval Metrics}

\textbf{Mean Reciprocal Rank (MRR):} Measures the average reciprocal rank of the first relevant chunk. Higher MRR indicates better ranking quality. Computed as: $MRR = \frac{1}{|Q|} \sum_{i=1}^{|Q|} \frac{1}{\text{rank}_i}$

\textbf{Hit Rate@k:} Binary metric measuring whether any relevant chunk appears in top-k results. We report Hit Rate@1 and Hit Rate@3.

\subsubsection{Domain-Specific Metrics}

\textbf{Citation Accuracy (UAE Legal only):} For legal domain, we manually evaluated whether cited sources were accurate and complete. Three annotators evaluated a random sample of 100 responses with high inter-annotator agreement (Fleiss' $\kappa = 0.82$).

\subsubsection{Statistical Testing}

We used paired t-tests to assess significance of performance differences, with a $p < 0.05$ threshold. Results averaged over 3 runs.

\section{Results}

\subsection{Main Results}

\begin{table}[htbp]
\centering
\caption{Performance on HotpotQA (Multi-Hop Reasoning). CDTA beats all baselines across metrics. Contextual improves over Semantic but cannot solve cross-document synthesis. Results averaged over 3 runs, all improvements statistically significant ($p < 0.05$).}
\label{tab:hotpotqa_results}
\renewcommand{\arraystretch}{1.2}
\small
\begin{tabular}{@{}lcccccc@{}}
\toprule
\textbf{Method} & \textbf{Faithfulness} & \textbf{Ans. Rel.} & \textbf{Context Recall} & \textbf{Context Prec.} & \textbf{MRR} & \textbf{Hit@1} \\
\midrule
Fixed-Size & 0.64 & 0.66 & 0.72 & 0.58 & 0.51 & 0.42 \\
Recursive & 0.71 & 0.75 & 0.78 & 0.66 & 0.58 & 0.51 \\
Semantic & 0.78 & 0.82 & 0.82 & 0.73 & 0.64 & 0.56 \\
Contextual & 0.83 & 0.86 & 0.86 & 0.78 & 0.69 & 0.63 \\
\midrule
\textbf{CDTA} & \textbf{0.93} & \textbf{0.90} & \textbf{0.97} & \textbf{0.89} & \textbf{0.86} & \textbf{0.88} \\
\bottomrule
\end{tabular}
\end{table}

CDTA reached 0.93 faithfulness on HotpotQA \citep{yang2018hotpotqa}, beating Contextual by 12\%, Semantic by 19\%, Recursive by 31\%, and Fixed-Size by 45\% (all $p < 0.05$). Context recall of 0.97 shows CDTA retrieves nearly complete information, versus 0.86 for Contextual and 0.82 for Semantic. Context precision of 0.89 indicates CDTA retrieves predominantly relevant information with minimal noise, establishing high signal-to-noise ratio that enables accurate answer generation.

Contextual augmentation improved over Semantic chunking by 6\% (0.83 vs 0.78 faithfulness), validating the utility of document-level context in embeddings. However, this improvement pales compared to CDTA's 19\% gain over Semantic. The pattern is clear: within-document context helps, but cross-document synthesis is what multi-hop reasoning demands.

For ranking quality, CDTA achieved 0.86 MRR, meaning relevant chunks typically appear at rank 1.16, compared to rank 1.45 for Contextual and 1.56 for Semantic. Hit Rate@1 of 0.88 means 88\% of queries find a relevant chunk in the very first result, versus 63\% for Contextual and 56\% for Semantic.

\begin{figure}[htbp]
\centering
\includegraphics[width=0.8\textwidth]{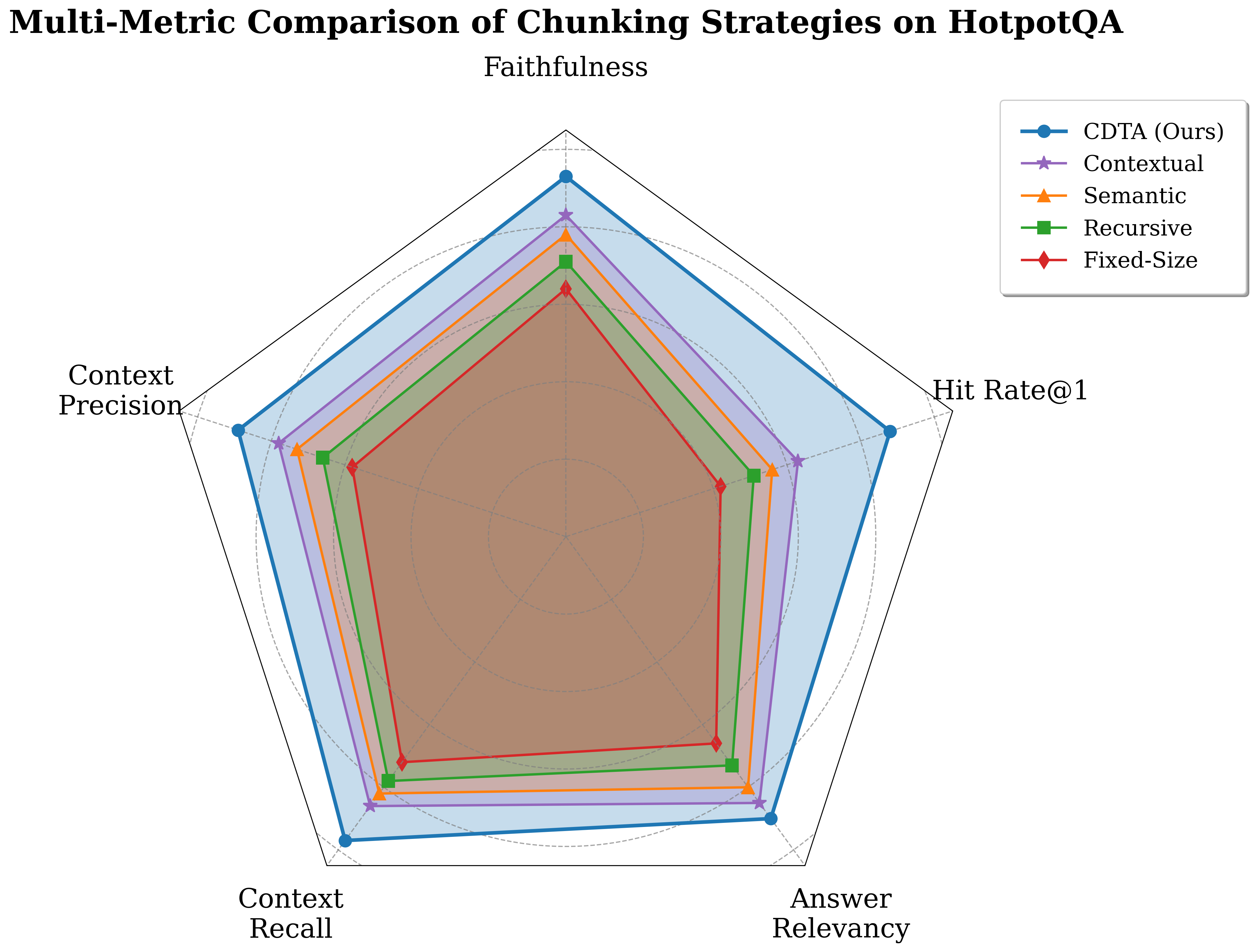}
\caption{Multi-metric comparison on HotpotQA. CDTA (blue, outer) versus Fixed-Size baseline (red, inner) shows superior performance across all dimensions: Faithfulness (0.93 vs 0.64), Hit Rate@1 (0.88 vs 0.42), Answer Relevancy (0.90 vs 0.66), Context Recall (0.97 vs 0.72), and Context Precision (0.89 vs 0.58).}
\label{fig:radar_comparison_final}
\end{figure}

\begin{table}[htbp]
\centering
\caption{Performance on UAE Legal dataset. CDTA outperforms all baselines across metrics. Contextual improves over Semantic but still fragments information. Results averaged over 3 runs ($p < 0.05$).}
\label{tab:legal_results}
\renewcommand{\arraystretch}{1.2}
\small
\begin{tabular}{@{}lcccc@{}}
\toprule
\textbf{Method} & \textbf{Faithfulness} & \textbf{Context Recall} & \textbf{Context Precision} & \textbf{Citation Accuracy} \\
\midrule
Fixed-Size & 0.58 & 0.68 & 0.53 & 0.52 \\
Recursive & 0.67 & 0.74 & 0.62 & 0.61 \\
Semantic & 0.74 & 0.79 & 0.71 & 0.68 \\
Contextual & 0.80 & 0.84 & 0.76 & 0.74 \\
\midrule
\textbf{CDTA} & \textbf{0.94} & \textbf{0.98} & \textbf{0.91} & \textbf{0.93} \\
\bottomrule
\end{tabular}
\end{table}

On domain-specific legal documents, CDTA reached 0.94 faithfulness and 0.93 citation accuracy, an 18\% improvement over Contextual, 27\% over Semantic, 40\% over Recursive, and 62\% over Fixed-Size (all $p < 0.05$). Context recall of 0.98 approaches perfect retrieval of necessary information, combined with 0.91 context precision. 

Contextual improved over Semantic by 8\% in this domain (0.80 vs 0.74), a larger gain than on HotpotQA, likely because legal documents have richer document-level structure that context summaries can capture. However, even with this advantage, Contextual still lags CDTA by 18\%, showing that document context cannot substitute for cross-document synthesis in domains where information systematically distributes across multiple authoritative sources.

Figure~\ref{fig:radar_comparison_final} illustrates CDTA's advantages across all evaluation dimensions relative to traditional fixed-size chunking.

\subsection{Retrieval Efficiency at Low k}

A key advantage emerges when examining performance under retrieval constraints. We analyzed faithfulness at $k=3$ (retrieving only 3 chunks) versus standard $k=5$:

\begin{table}[htbp]
\centering
\caption{Faithfulness at $k=3$ vs $k=5$ on HotpotQA. CDTA maintains high performance with fewer chunks, showing information density advantages. Results averaged over 3 runs ($p < 0.05$).}
\label{tab:low_k}
\renewcommand{\arraystretch}{1.2}
\begin{tabular}{@{}lccc@{}}
\toprule
\textbf{Method} & \textbf{k=3} & \textbf{k=5} & \textbf{Degradation} \\
\midrule
Fixed-Size & 0.51 & 0.64 & -20\% \\
Recursive & 0.59 & 0.71 & -17\% \\
Semantic & 0.68 & 0.78 & -13\% \\
Contextual & 0.74 & 0.83 & -11\% \\
\midrule
\textbf{CDTA} & \textbf{0.91} & \textbf{0.93} & \textbf{-2\%} \\
\bottomrule
\end{tabular}
\end{table}

\begin{figure}[htbp]
\centering
\includegraphics[width=0.9\textwidth]{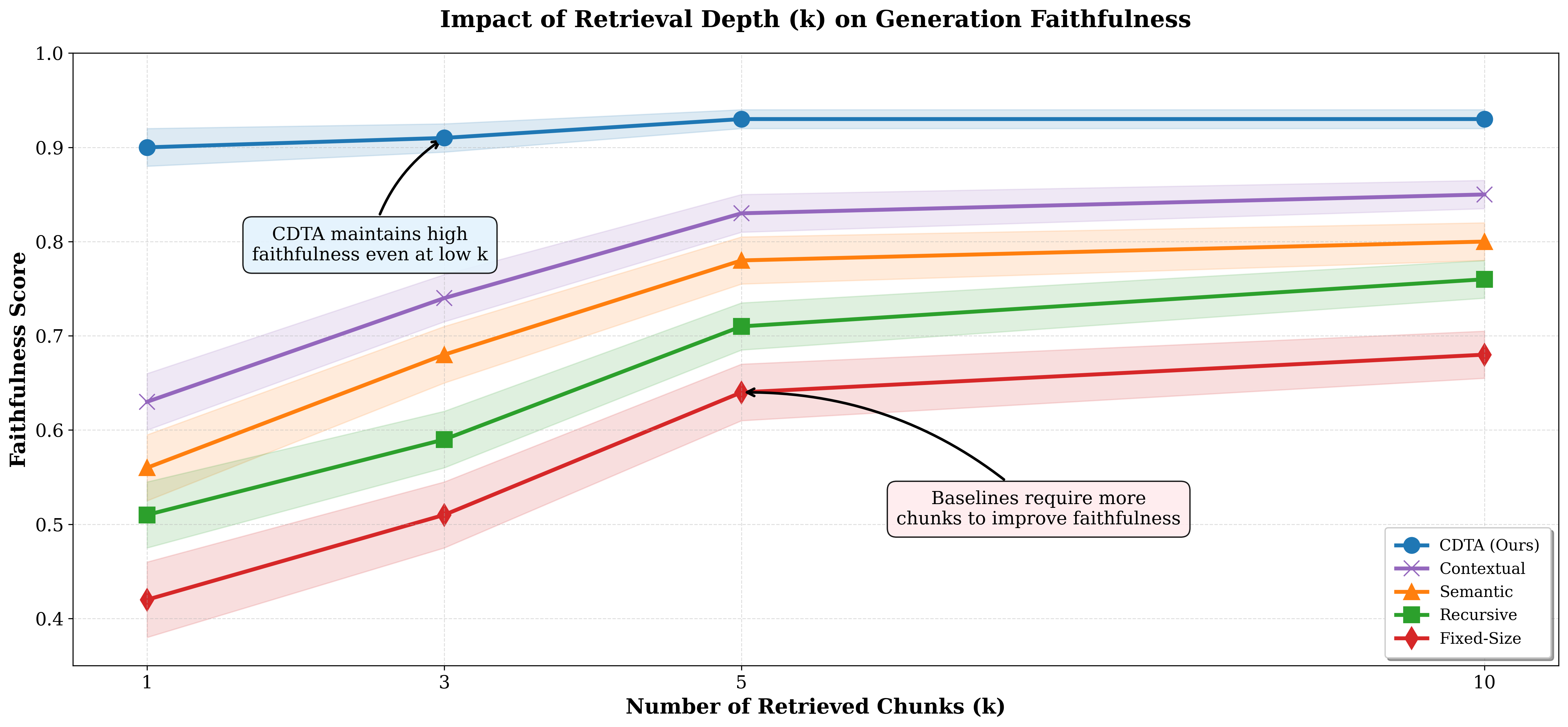}
\caption{Faithfulness on HotpotQA as $k$ increases from 1 to 10. CDTA maintains consistently high performance (0.90-0.93) across all $k$ values, while traditional methods require more chunks to achieve competitive results. Error bands represent standard deviation across 3 runs.}
\label{fig:performance_vs_k}
\end{figure}

CDTA suffers only 2\% faithfulness degradation when reducing $k$ from 5 to 3, compared to 11\% for Contextual, 13\% for Semantic, 17\% for Recursive, and 20\% for Fixed-Size. This resilience stems from producing information-dense chunks where a single retrieved unit contains comprehensive topic coverage. Contextual performs better than Semantic here (11\% vs 13\% degradation) because document context helps identify the most relevant chunks, but still requires retrieving multiple fragments to assemble complete information.

Figure~\ref{fig:performance_vs_k} shows CDTA's remarkable resilience at low $k$ values, maintaining high faithfulness even when retrieving minimal context. Contextual shows modest improvement over Semantic at all $k$ values, but both require substantially larger $k$ to approach CDTA's performance.

\begin{table}[htbp]
\centering
\caption{Hit Rate@k on HotpotQA. CDTA achieves higher rates of relevant chunk retrieval in top positions. Results averaged over 3 runs ($p < 0.05$).}
\label{tab:hit_rate}
\renewcommand{\arraystretch}{1.2}
\begin{tabular}{@{}lcc@{}}
\toprule
\textbf{Method} & \textbf{Hit Rate@1} & \textbf{Hit Rate@3} \\
\midrule
Fixed-Size & 0.42 & 0.68 \\
Recursive & 0.51 & 0.76 \\
Semantic & 0.56 & 0.79 \\
Contextual & 0.63 & 0.84 \\
\midrule
\textbf{CDTA} & \textbf{0.88} & \textbf{0.96} \\
\bottomrule
\end{tabular}
\end{table}

CDTA achieves 0.88 Hit Rate@1, 88\% of queries find a relevant chunk in the very first result, compared to 0.63 for Contextual, 0.56 for Semantic, 0.51 for Recursive, and 0.42 for Fixed-Size. At $k=3$, CDTA reaches 0.96, ensuring relevant information appears in the top three results for nearly all queries. Contextual's Hit Rate@1 of 0.63 beats Semantic's 0.56, confirming that document context improves retrieval ranking. However, this 7-point gain is dwarfed by CDTA's 32-point advantage over Semantic (0.88 vs 0.56).

\subsection{Query-Type Analysis}

Analyzing HotpotQA results by question type revealed differential performance patterns:

\begin{table}[htbp]
\centering
\caption{Faithfulness on HotpotQA by query type. CDTA's advantage concentrates in multi-document queries, where Contextual also improves over Semantic but still lags. Results averaged over 3 runs ($p < 0.05$).}
\label{tab:query_type}
\renewcommand{\arraystretch}{1.2}
\small
\begin{tabular}{@{}lcccccc@{}}
\toprule
\textbf{Query Type} & \textbf{Fixed} & \textbf{Recursive} & \textbf{Semantic} & \textbf{Contextual} & \textbf{CDTA} & \textbf{Gain vs Ctx} \\
\midrule
Bridge (2 docs) & 0.61 & 0.69 & 0.76 & 0.81 & \textbf{0.92} & +14\% \\
Comparison (2+ docs) & 0.64 & 0.72 & 0.79 & 0.84 & \textbf{0.95} & +13\% \\
Simple (1 doc) & 0.71 & 0.79 & 0.82 & 0.85 & \textbf{0.93} & +9\% \\
\bottomrule
\end{tabular}
\end{table}

CDTA's advantage concentrates in query types requiring cross-document synthesis (bridge and comparison questions) with 13-14\% gains over Contextual, versus 9\% gain for simple factual queries. Contextual improves over Semantic across all query types (5-6\% gains), showing document context helps even for single-document queries. However, Contextual's gains remain modest compared to CDTA's, particularly for multi-document questions where information fragmentation is most acute.

This pattern validates our hypothesis: cross-document synthesis benefits queries that inherently require it most. Contextual helps by providing document-level framing, but cannot substitute for actual information consolidation across sources.

\subsection{Efficiency Considerations}

\textbf{Indexing Costs.} CDTA's indexing process is more expensive than traditional methods. For a 1,000-document corpus:
\begin{itemize}[noitemsep]
\item Fixed-Size: 0.1 seconds per document
\item Semantic: 2-3 seconds per document
\item Contextual: 4-6 seconds per document (semantic + context generation)
\item \textbf{CDTA: 38-44 seconds per document}
\end{itemize}

This cost is incurred only once during indexing. For HotpotQA (5,901 documents), total indexing time was approximately 32 hours on our infrastructure (parallel processing with 8 workers). API costs for the UAE Legal corpus (847 documents) totaled approximately \$420 for CDTA (\$180 topic extraction, \$195 relevance mapping, \$45 synthesis) versus \$78 for Contextual (context generation only).

\textbf{Query Performance.} The information density of CDTA chunks enables efficient query processing. Retrieving fewer chunks to obtain necessary information reduces generator token consumption and speeds response generation. In our experiments, average query latency was 2.3s for CDTA vs 2.6s for Contextual vs 2.7s for Semantic ($k=5$).

\textbf{Break-Even Analysis.} For applications with high query volumes (more then 10,000 queries per month), improved retrieval efficiency can offset higher indexing costs through reduced compute costs. The exact break-even point depends on specific deployment costs, query volumes, and infrastructure, but for typical RAG deployments with thousands of queries daily, the improved efficiency of CDTA chunks can justify the upfront indexing investment within several months of operation.

\subsection{Qualitative Analysis}

\begin{mdframed}[
  frametitle={Qualitative Comparison: UAE Legal Data Retention Query},
  frametitlebackgroundcolor=boxedge,
  frametitlefontcolor=black,
  backgroundcolor=boxback,
  roundcorner=5pt,
  linewidth=1pt,
  linecolor=boxedge,
  skipabove=\baselineskip,
  skipbelow=\baselineskip
]
\textbf{Query:} ``What are the data retention requirements for financial records?''

\vspace{0.5em}
\hrule
\vspace{0.5em}

\textbf{Semantic Chunking} retrieved three disconnected fragments: base six-year requirement, vague UAE national exception, and mention of a 2019 change. Generated answer was incomplete, stating only ``six years, with some exceptions for UAE nationals.''

\vspace{0.5em}
\hrule
\vspace{0.5em}

\textbf{Contextual Retrieval} added document labels (DIFC Law 2007, Federal Decree 2018, DIFC Amendment 2019) to the same fragments. This improved attribution but didn't solve fragmentation. Generated answer cited sources but hedged: ``six years under DIFC Law 2007, with some exceptions per Federal Decree 2018. This may have changed in 2019.''

\vspace{0.5em}
\hrule
\vspace{0.5em}

\textbf{CDTA Retrieval} provided a single synthesized chunk titled ``Data Retention Requirements - Financial Institutions'' containing: base requirement (six years, DIFC Law No. 1/2007), 2019 amendment (extended to eight years), UAE national exception (ten years, Federal guidance), scope expansion (electronic communications, 2022 court ruling), and approved storage jurisdictions. Generated answer was complete and accurate: ``Financial institutions must retain customer records for eight years under DIFC Law as amended in 2019, with ten-year requirement for UAE nationals per Federal Decree. Records include all electronic communications per DIFC Court ruling 2022-CFI-045. Storage permitted in approved jurisdictions per regulatory guidance.''
\end{mdframed}

The comparison illustrates how cross-document synthesis solves the signal-to-noise problem: traditional approaches delivered low-signal fragments, Contextual added document labels to those fragments, while CDTA provided high-signal, low-noise context making answer generation straightforward.

\section{Ablation Studies}

To understand which components contribute most to performance, we conducted ablation experiments on HotpotQA:

\begin{table}[htbp]
\centering
\caption{Ablation study on HotpotQA. Both relevance mapping and LLM-driven synthesis are critical. Results averaged over 3 runs ($p < 0.05$).}
\label{tab:ablation}
\renewcommand{\arraystretch}{1.2}
\begin{tabular}{@{}llcl@{}}
\toprule
\textbf{Configuration} & \textbf{Faithfulness} & \textbf{Change} & \textbf{Description} \\
\midrule
\textbf{Full CDTA} & \textbf{0.93} & Baseline & Complete 6-stage pipeline \\
Without Synthesis & 0.82 & -11\% & Stage 5 removed (concatenation) \\
Without Relevance Filter & 0.87 & -6\% & Stages 3A/3B removed \\
\bottomrule
\end{tabular}
\end{table}

\textbf{Synthesis is essential.} Replacing LLM-driven synthesis with simple concatenation of relevant segments resulted in 11-point degradation ($p < 0.05$). Coherent reconstruction rather than mere aggregation drives performance gains.

\textbf{Relevance mapping is critical.} Removing the relevance filtering stage caused 6-point faithfulness drop ($p < 0.05$). Without filtering, topic aggregations included tangentially related segments that introduced noise, degrading the quality.

\section{Discussion}

Results show our approach offers performance advantages for multi-hop reasoning, validating our core hypothesis. However, the value of this computationally intensive approach depends on specific deployment contexts \citep{zhao2024retrieval}. We now analyze conditions under which this new method excels and scenarios where simpler methods may suffice.

\subsection{When CDTA Excels}

It shows advantages in information environments characterized by knowledge fragmentation, common in specialized domains.

\textbf{High Information Fragmentation.} CDTA is purpose-built for domains where knowledge about single topics systematically distributes across multiple documents. This includes regulatory compliance, where complete understanding requires synthesizing foundational laws, amendments, and guidance notes, a core challenge of legal reasoning and knowledge fragmentation well-documented in the field \citep{zhong2020does,henderson2022pile,chalkidis2019deep}.

\textbf{Temporal Knowledge Evolution.} The methodology excels for corpora where information evolves through separately published updates. Document-centric approaches, even with contextual augmentation, cannot provide single representations of current law, whereas CDTA synthesizes base legal statutes with later modifications.

\textbf{Complex, Multi-Hop Queries.} As HotpotQA results show \citep{yang2018hotpotqa}, CDTA's performance advantage concentrates in queries inherently requiring fact combination from multiple sources. For applications focused on complex question answering, automated literature reviews, or in-depth financial analysis, pre-synthesis of topic knowledge directly aligns knowledge base structure with query nature.

\subsection{When Simpler Methods Suffice}

CDTA's indexing overhead offers diminishing returns in scenarios where the problem is less prevalent.

\textbf{Single-Document Queries.} For knowledge bases consisting of largely independent, encyclopedic articles where answers are typically self-contained within single documents, contextual augmentation offers a more efficient solution. Our results show Contextual improves over Semantic by 3-4\% for single-document queries while requiring only 2-3x indexing cost versus CDTA's 10x cost.

\textbf{Cost-Constrained or Real-Time Indexing.} For applications with extremely limited indexing budgets or requiring near-instantaneous indexing of new documents, CDTA's computational cost and latency (38-44s per document) may be prohibitive. Contextual augmentation (4-6s per document) offers a middle ground, better than simple methods while remaining tractable.

\subsection{Industry Methods as Baselines}

We included Contextual Retrieval as a baseline despite its origin in industry engineering documentation rather than peer-reviewed literature. This decision reflects a key reality in RAG system development: foundational techniques often emerge from industry labs and are deployed in production before academic publication. Contextual Retrieval now serves as best practice recommended by leading AI companies and actively used in real-world systems.

For a benchmark to be meaningful for practitioners, it must compare against methods they're actually considering. Ignoring industry techniques would make our results less relevant to real-world deployment decisions. However, we maintain scientific rigor by implementing and testing Contextual under the same controlled conditions as academic baselines, identical datasets, metrics, and evaluation protocols.

\subsection{Deployment Considerations}

Successfully transitioning CDTA to production requires addressing several practical challenges.

\textbf{Managing Cost and Latency.} The primary operational challenge is the cost of LLM-intensive indexing pipeline. Mitigation strategies include using tiered model architecture, smaller, faster models (like GPT-4o-mini) for high-volume tasks like binary relevance checks, while reserving powerful models (like Claude 3.5 Sonnet) for nuanced synthesis. Process optimization through batching and parallelizing LLM API calls maximizes throughput.

\textbf{Quality Control.} Final knowledge base quality directly depends on LLM performance. Prompts for topic extraction, relevance mapping, and synthesis are critical infrastructure requiring iterative testing against diverse document types. Post-synthesis splitting strategies handle cases where synthesis generates very long topic chunks exceeding optimal sizes for downstream retrieval.

\textbf{Update Strategies.} This method is best suited for relatively static corpora. For corpora with periodic updates, implementing targeted re-computation is more efficient than full re-indexing. When a source document is modified, the system can identify all synthesized topic chunks that cited that document and trigger re-synthesis only for that specific subset.

\textbf{Ethical Considerations.} Deploying in production requires careful attention to potential risks. LLMs can amplify biases present in source documents during synthesis, organizations should conduct bias audits on synthesized content, particularly for sensitive domains like legal or medical applications. While CDTA maintains strict fidelity to source material, synthesis could inadvertently create misleading connections between facts. Citation metadata and traceability to source documents are essential safeguards. In domains like legal and medical information, errors in synthesized content could have serious consequences, human expert review of high-stakes content is advisable before deployment.

\subsection{Limitations}

CDTA has several limitations worth noting:

\textbf{Computational Cost.} High indexing cost (\$420 for 847 documents versus \$78 for Contextual) is the primary barrier to adoption. This makes CDTA most practical for high-value, relatively static knowledge bases where cost can be amortized over high query volumes.

\textbf{LLM Dependency.} Entire pipeline quality depends on underlying LLM capabilities and potential biases. Errors in topic extraction or relevance mapping can propagate and degrade final synthesized chunk quality.

\textbf{Scalability.} The $O(|T| \times |S|)$ complexity of relevance mapping remains challenging for corpora at the scale of millions of documents. For our 847-document corpus, this required 756,000 API calls. Further optimization using advanced pre-filtering techniques is necessary for larger deployments.

\subsection{Future Directions}

Several promising research directions could extend this work:

\textbf{Hybrid Models.} Combining CDTA with hierarchical intra-document methods like RAPTOR \citep{sarthi2024raptor} could take the best from both approaches. CDTA could first perform cross-document synthesis to generate comprehensive topic documents, then RAPTOR could be applied to each synthesized document to create multi-level, searchable tree structures. Similarly, it could be combined with contextual augmentation, first unifying distributed information across documents through CDTA, then applying contextual techniques to preserve document-specific context within each synthesized chunk.

\textbf{Dynamic Synthesis.} To mitigate high upfront indexing cost, a ``late synthesis'' approach could perform LLM-driven synthesis at query time rather than indexing time. The system would retrieve relevant base segments from multiple documents and synthesize on-the-fly. While this would increase query latency, it would eliminate expensive indexing phases and better suit highly dynamic data.

\textbf{Specialized Models.} Cost is largely driven by reliance on large, general-purpose LLMs. Development and fine-tuning of smaller, specialized models explicitly trained for multi-source consolidation, de-duplication, and coherent summarization could potentially achieve comparable quality at a fraction of the cost.

\section{Conclusion}

Performance of RAG systems is determined by chunking yet it remains understudied. While the field has made strides in preserving context within documents, knowledge fragmentation across remains a fundamental bottleneck for complex, multi-source queries. We introduce CDTA chunking, which reframes chunking from document partitioning to corpus-level knowledge reconstruction. By identifying topics globally and using LLMs to synthesize all relevant information into complete, coherent knowledge units, it directly tackles the distributed information challenge that has long challenged knowledge representation systems \citep{sowa2006challenge}.

Our evaluation shows improvements in generation quality, particularly for tasks requiring multi-document reasoning. On multi-hop question answering \citep{yang2018hotpotqa}, CDTA reached 0.93 faithfulness, a 19\% improvement over semantic chunking and 12\% over contextual augmentation. The intensive indexing process is a computing investment, but leads to lower query times and a clear deployment profile for applications where retrieval quality and speed matter most.

The trade-off between higher indexing costs and superior query-time performance positions CDTA as a tool for knowledge-intensive applications with relatively static corpora and high query volumes, such as regulatory compliance, technical support, and medical or legal analysis. By structuring information around topics, mirroring human conceptual organization rather than arbitrary textual boundaries, we enable more precise understanding. Future work will focus on optimizing the indexing pipeline and exploring hybrid models to broaden the applicability of this knowledge reconstruction paradigm.

\clearpage
\bibliographystyle{plainnat}
\bibliography{references}

\clearpage
\appendix
\section*{Appendix A: Prompt Templates}
\label{sec:appendix_prompts}

\subsection*{A.1 Topic Extraction Prompt}
\label{subsec:topic_extraction}

\begin{promptbox}
\small
\textbf{Role:} You are a meticulous AI research assistant specializing in semantic analysis.

\textbf{Objective:} Analyze the provided text segment and identify all distinct, high-level conceptual topics it discusses. For each topic, generate a focused description and a one-sentence summary.

\textbf{Instructions:}
\begin{enumerate}[noitemsep]
\item Read the entire segment carefully to understand its content.
\item Identify conceptual themes, not just keywords. For example, a discussion about ``server uptime requirements'' and ``disaster recovery protocols'' should be identified as a topic like ``System Reliability and Business Continuity.''
\item Generate a unique name for each topic that is a noun phrase.
\item Generate a description that crisply summarizes the topic's scope.
\item Format your response as valid JSON containing a list under the key ``topics''.
\end{enumerate}

\textbf{Text Segment:}
\textit{[Document Segment Text]}

\textbf{Response Format:}
\begin{verbatim}
{
  "topics": [
    {
      "name": "[Topic Name]",
      "description": "[Brief description]"
    }
  ]
}
\end{verbatim}
\end{promptbox}

\clearpage
\subsection*{A.2 Relevance Check Prompt}
\label{subsec:relevance_mapping}

\begin{promptbox}
\small
\textbf{Role:} You are an AI relevance classifier making strict binary judgments.

\textbf{Topic Name:} \textit{[Topic Name]}

\textbf{Topic Description:} \textit{[Topic Description]}

\textbf{Text Segment:} \textit{[Document Segment Text]}

\textbf{Question:} Does the text segment above contain information that is directly and substantively relevant to the topic?

\textbf{Respond with only a single word:} ``true'' or ``false''.
\end{promptbox}

\clearpage
\subsection*{A.3 Knowledge Synthesis Prompt}
\label{subsec:synthesis_prompt}

\begin{promptbox}
\small
\textbf{Role:} You are an expert AI knowledge synthesizer.

\textbf{Objective:} Synthesize the provided source fragments into a single, well-structured, coherent document that comprehensively explains the topic.

\textbf{Topic:} \textit{[Topic Name]}

\textbf{Topic Description:} \textit{[Topic Description]}

\textbf{Source Fragments:}
\begin{itemize}[noitemsep]
\item ``\textit{[Document Chunk 1]}''
\item ``\textit{[Document Chunk 2]}''
\item ...
\end{itemize}

\textbf{Instructions:}
\begin{enumerate}[noitemsep]
\item \textbf{Comprehensiveness:} Include every piece of relevant information from the sources.
\item \textbf{Coherence:} Weave information into a logical narrative with clear structure.
\item \textbf{Fidelity:} Adhere strictly to source information. Do not add external knowledge. If sources conflict, note the contradiction.
\item \textbf{Conciseness:} Remove redundancy. If multiple fragments state the same fact, present it once.
\item \textbf{Attribution:} Note which sources contributed key information where appropriate.
\end{enumerate}

\textbf{Respond only with the synthesized document text.}
\end{promptbox}

\end{document}